\documentclass{DISproc}

\usepackage{relsize}
\usepackage{amsmath}
\usepackage{wrapfig}  
\usepackage{amssymb} 
\def\babar{\mbox{\slshape B\kern-0.1em{\smaller A}\kern-0.1em B\kern-0.1em{\smaller A\kern-0.2em R }}} %

\begin{document}
%------------------------------------
\title{Charmonium and charmonium-like results from \babar}
\author{{\slshape Elisa Fioravanti$^1$}\\[1ex]
$^1$INFN Ferrara, Via Saragat 1, 44122, Ferrara, Italy\\
}

% please enter the contribution ID for the DOI
\contribID{xy}

\doi  % if there is an online version we will register DOIs

\maketitle

\begin{abstract}
  We present new results on charmonium and charmonium-like states from the \babar experiment located at the
PEP-II asymmetric energy $e^+e^-$ storage ring at the SLAC National Accelerator Laboratory.
  \end{abstract}

\section {Study of the process $\gamma\gamma\rightarrow J/\psi\omega$}
\begin{figure}[htb]
  \centering
  \includegraphics[width=0.55\textwidth]{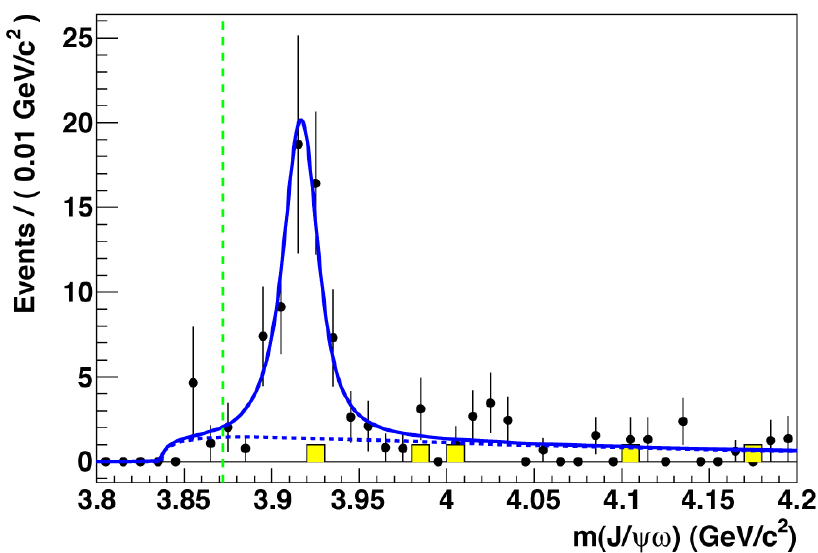}
  \caption{The efficiency-corrected invariant mass distribution for the $J/\psi\omega$ final state. The vertical dashed line is placed at the X(3872) mass.}
  \label{Fig:Xres1}
\end{figure}

The charmonium-like state X(3915) was first observed by Belle \cite{belleX3915} in two-photon fusion events decaying into $J/\psi\omega$. In addition, it was seen decaying into $J/\psi\omega$ in B decays by \babar \cite{babarX3915B}, along with the X(3872).\\
We study the process $\gamma\gamma\rightarrow J/\psi\omega$ at \babar to search for the X(3915) and the X(3872) resonances via the decay to $J/\psi\omega$, using a data sample of 519 fb$^{-1}$. Figure \ref{Fig:Xres1} presents the reconstructed $J/\psi\omega$ invariant mass distribution after all the selection criteria have been applied. We perform an extended maximum likelihood fit to the efficiency-corrected spectrum.  A large peak at near 3915 MeV/c$^2$ is observed with a significance of 7.6$\sigma$. The measured resonance parameters are $m[X(3915)]=(3919.4\pm2.2\pm1.6)$ MeV/c$^2$, $\Gamma[X(3915)]=(13\pm6\pm3)$ MeV. The measured value of the two-photon width times the branching fraction, $\Gamma_{\gamma\gamma}[X(3915)]$ x $\cal{B}$$(X(3915)\rightarrow J/\psi\omega)$ is ($52\pm10\pm3)$ eV and $(10.5\pm1.9\pm0.6)$ eV for two spin hypotheses $J=0$ and $J=2$, respectively, where the first error is statistical and the second is systematic. In addition, a Bayesian upper limit (UL) at 90\% confidence level (CL) is obtained for the X(3872), $\Gamma_{\gamma\gamma}[X(3872)]$ x $\cal{B}$$(X(3872)\rightarrow J/\psi\omega)<$ 1.7 eV, assuming J=2.

\section {Study of the process $\gamma\gamma\rightarrow\eta_c\pi^+\pi^-$}
\begin{figure}[htb]
  \centering
  \includegraphics[width=0.75\textwidth]{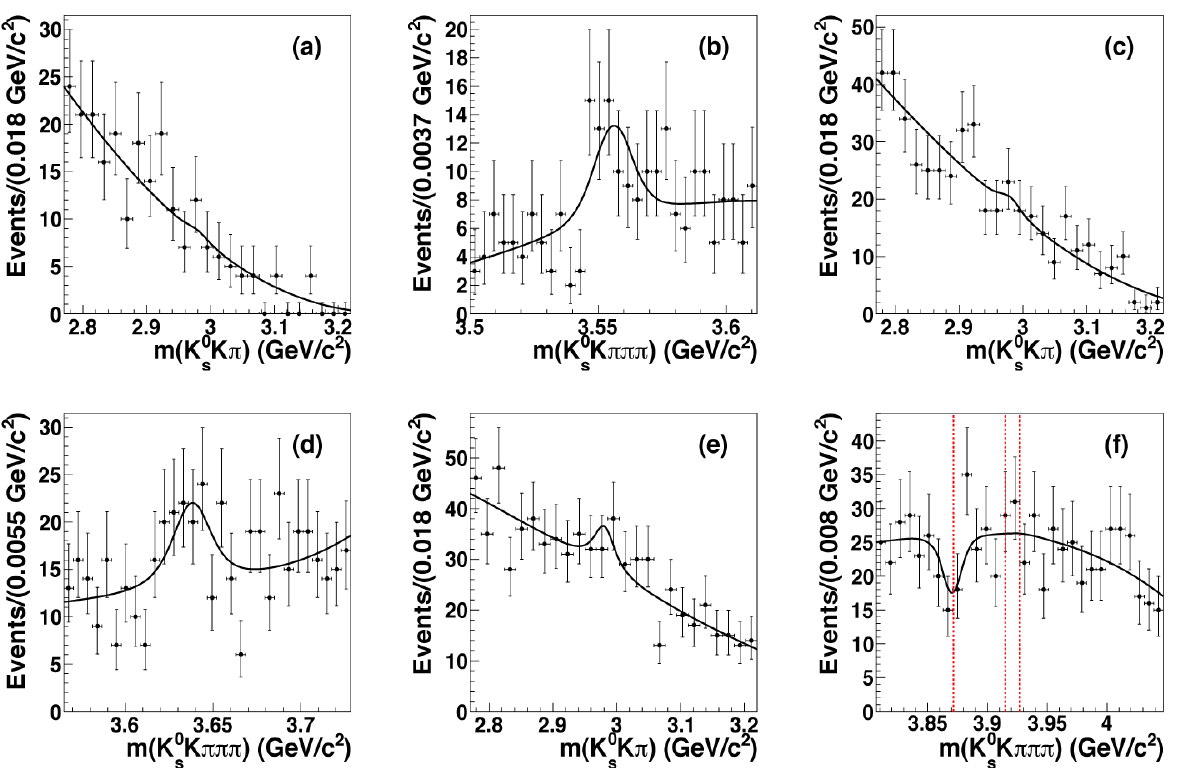}
  \caption{Distributions of (a,c,e) $m(K_S^0K^+\pi^-)$ and (b,d,f)  $m(K_S^0K^+\pi^-\pi^+\pi^-)$ with the fit function overlaid for the fit regions of the (a,b) $\chi_{c2}(1P)$, (c,d) $\eta_c(2S)$, and (e,f) X(3872), X(3915) and $\chi_{c2}(2P)$. The vertical dashed lines in (f) indicates the peak mass positions of the X(3872), X(3915) and $\chi_{c2}(2P)$.}
  \label{Fig:etapipi}
\end{figure}

This analysis has been studied for the first time and is performed to search for resonances decaying into $\eta_c\pi^+\pi^-$, using a data sample of 474 fb$^{-1}$. The $\eta_c$ was reconstructed via its decay to $K_S^0K^+\pi^-$, with $K_S^0\rightarrow\pi^+\pi^-$. The signal yield for each X resonance is extracted from a two-dimensional fit to $m(K_S^0K^+\pi^-)$  and $m(K_S^0K^+\pi^-\pi^+\pi^-)$. Figure \ref{Fig:etapipi} presents the two dimensional fits around each of the resonances. No significant signal is observed in any of the fits. Table \ref{Tab:tabetapipi} summarizes these results. ULs are obtained on the branching fractions $\cal{B}$$(\eta_c(2S)\rightarrow\eta_c\pi^+\pi^-)<$ 7.4\% and $\cal{B}$$(\chi_{c2}(1P)\rightarrow\eta_c\pi^+\pi^-)<$ 2.2\% at 90\% CL. 

\begin{table}[!htbp]
\centering
\begin{tabular}{|c|c|c|c|c|}
\hline
Resonances & $M_X$ (MeV/c$^2$) & $\Gamma_X$ (MeV) &  \multicolumn{2}{c|}{$\Gamma_{\gamma\gamma}\cal{B}$ (eV)}\\
& & & Central Value & UL\\
\hline
$\chi_{c2}(1P)$  &  3556.20$\pm$0.09 & 1.97$\pm$0.11 & $7.2^{+5.5}_{-4.4}\pm2.9$ & 15.7\\
$\eta_c(2S)$ & 3638.5$\pm$1.7 & 13.4$\pm$5.6 & $65^{+47}_{-44}\pm18$ & 133\\
$X(3872)$ & 3871.57$\pm$0.25 & 3.0$\pm$2.1 & $-4.5^{+7.7}_{-6.7}\pm2.9$ & 11.1\\
$X(3915)$ & 3915.0$\pm$3.6 & 17.0$\pm$10.4 & $-13^{+12}_{-12}\pm8$ & 16\\
$\chi_{c2}(2P)$ & 3927.2$\pm$2.6 & 24$\pm$6 & $-16^{+15}_{-14}\pm6$ & 19\\
\hline
\end{tabular}
\caption{Results of the $\gamma\gamma\rightarrow\eta_c\pi^+\pi^-$ fits. For each resonance X, we show the peak mass and width used in the fit; the product of the two-photon partial width $\Gamma_{\gamma\gamma}$ and the $X\rightarrow\eta_c\pi\pi$ branching fraction, and the 90\% CL upper limits on this product. }
\label{Tab:tabetapipi}
\end{table}

\section {Search for the $Z_1(4050)^+$ and $Z_2(4250)^+$}
\begin{figure}[htb]
  \centering
  \includegraphics[width=1.0\textwidth]{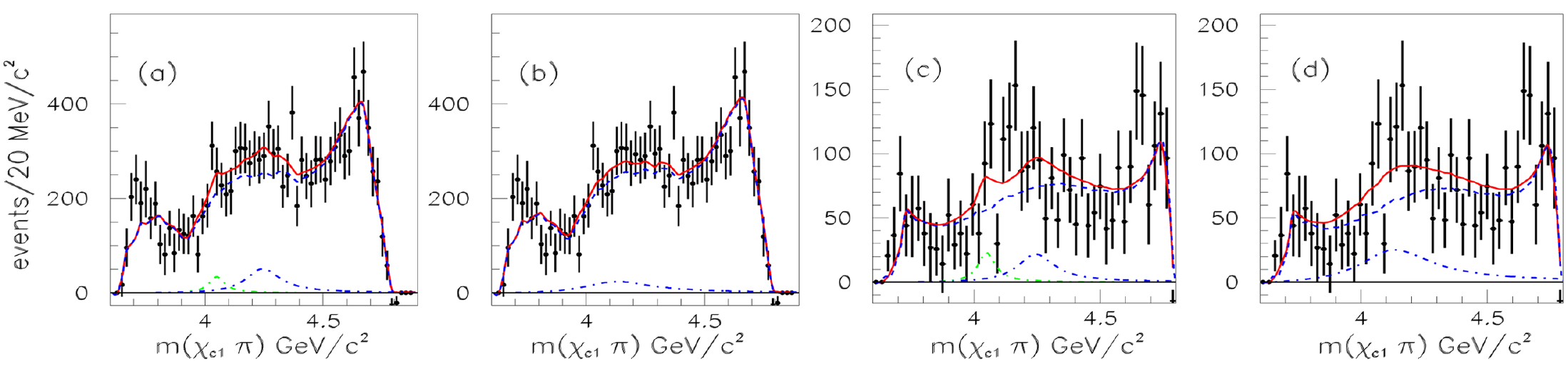}
  \caption{Fit on the background-subtracted and efficiency-corrected $\chi_{c1}\pi$ mass distribution. See text for more details.}
  \label{Fig:antimo}
\end{figure}

Belle reported the observation of two resonance-like structures, $Z_1(4050)^+$ and $Z_2(4250)^+$ in the study of $\bar{B}^0\rightarrow\chi_{c1}K^-\pi^+$, both decaying to $\chi_{c1}\pi^+$ \cite{belleZ}. \\
\babar studied the same final states \cite{antimo} to search for the $Z_1(4050)^+$ and $Z_2(4250)^+$ decay into $\chi_{c1}\pi^+$ in $\bar{B}^0\rightarrow\chi_{c1}K^-\pi^+$ and $B^+\rightarrow K_S^0\chi_{c1}\pi^+$ where $\chi_{c1}\rightarrow J/\psi\gamma$, using a data sample of 429 fb$^{-1}$. The $\chi_{c1}\pi^+$ mass distribution, background-subtracted and efficiency-corrected, was modeled using the $K\pi$ mass distribution and the corresponding normalized $K\pi$ Legendre polynomial moments. Figure \ref{Fig:antimo} shows the results of the fits done on the $\chi_{c1}\pi^+$ mass spectrum. The fit shown in Figure \ref{Fig:antimo}(a) includes both $Z_1(4050)^+$ and $Z_2(4250)^+$ resonances and the fit shown in Figure \ref{Fig:antimo}(b) includes a single broad $Z(4150)^+$ resonance. The Figures \ref{Fig:antimo}(c,d) show the $\chi_{c1}\pi$ mass spectrum fitted in the Dalitz plot region 1.0 $\le$ $m^2(K\pi)<$ 1.75 GeV$^2$/c$^4$ in order to make a direct comparison to the Belle results \cite{belleZ} (this region is labeled as "window" in Table \ref{Tab:antimo}). The results of the fits are summarized in Table \ref{Tab:antimo} and in every case the yield significance does not exceed 2$\sigma$. The ULs on the 90\% CL on the branching fractions are: $\cal{B}$$(\bar{B}^0\rightarrow Z_1(4050)^+K^-)$ x $\cal{B}$$(Z_1(4050)^+\rightarrow\chi_{c1}\pi^+)<$ 1.8 x 10$^{-5}$; $\cal{B}$$(\bar{B}^0\rightarrow Z_2(4250)^+K^-)$ x $\cal{B}$$(Z_2(4250)^+\rightarrow\chi_{c1}\pi^+)<$ 4.0 x 10$^{-5}$ and $\cal{B}$$(\bar{B}^0\rightarrow Z^+K^-)$ x $\cal{B}$$(Z^+\rightarrow\chi_{c1}\pi^+)<$ 4.7 x 10$^{-5}$.

\begin{table}[!htbp]
\centering
\begin{tabular}{|c|c|c|c|c|}
\hline
Data & Resonances & $N_{\sigma}$ & Fraction (\%) & $\chi^2/NDF$ \\
\hline
a) Total & $Z_1(4050)^+$ & 1.1 & 1.6$\pm$1.4 & 57/57\\
              & $Z_2(4250)^+$ & 2.0 & 4.8$\pm$2.4 & \\
b) Total & $Z(4150)^+$ & 1.1 & 4.0$\pm$3.8 & 61/58\\
\hline
a) Window & $Z_1(4050)^+$ & 1.2 & 3.5$\pm$3.0 & 53/46\\
              & $Z_2(4250)^+$ & 1.3 & 6.7$\pm$5.1 & \\
b) Window & $Z(4150)^+$ & 1.7 & 1.37$\pm$8.0 & 53/47\\
\hline
\end{tabular}
\caption{Results of the $\chi_{c1}\pi$ fits. $N_\sigma$ and Fraction give, for each fit, the significance and the fractional contribution of the Z resonances.}
\label{Tab:antimo}
\end{table}

\section {Study of the $J/\psi\pi^+\pi^-$ via Initial State Radiation (ISR)}
\begin{figure}[htb]
  \centering
  \includegraphics[width=0.85\textwidth]{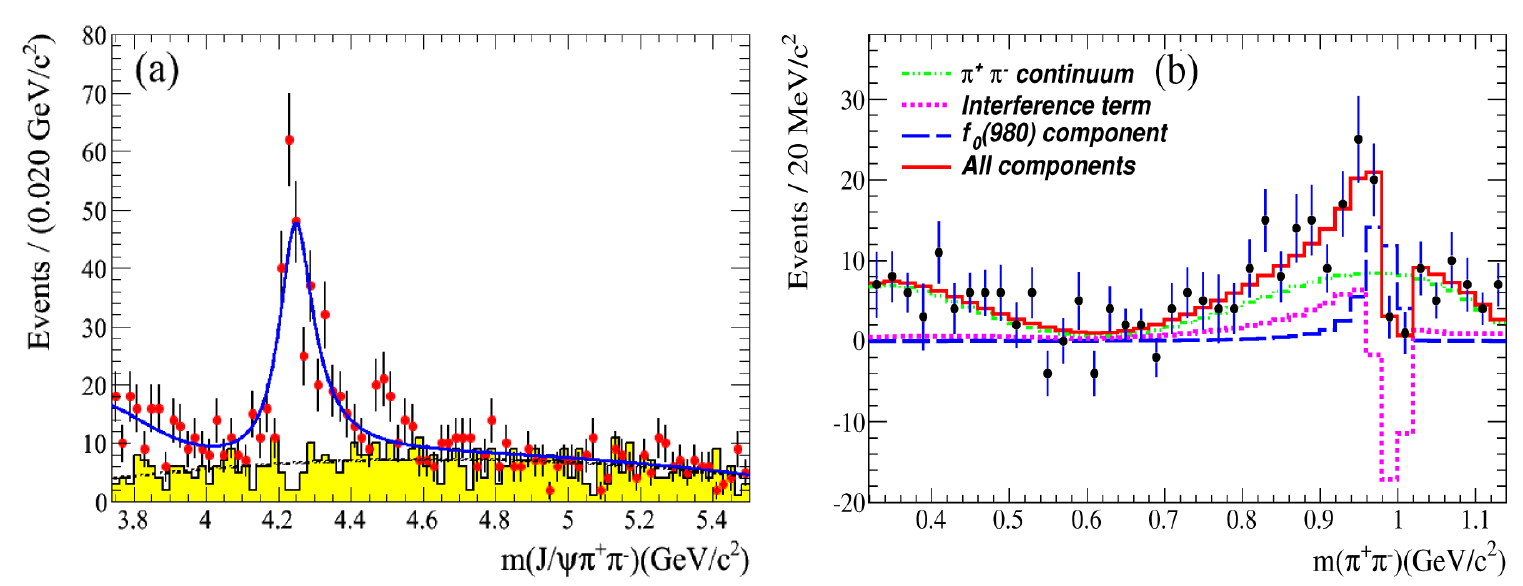}
  \caption{Figure (a) shows the fit to the $J/\psi\pi^+\pi^-$ invariant mass distribution. The Figure (b) shows the fit to the dipion invariant mass distribution.}
  \label{Fig:vale1}
\end{figure}

The Y(4260) charmonium-like resonance was discovered by \babar \cite{Y} in ISR production of $J/\psi\pi^+\pi^-$. A subsequent Belle analysis \cite{Ybelle} of the same final state suggested also the existence of an additional resonance around 4.1 GeV/c$^2$ that they dubbed the Y(4008).\\
This analysis \cite{valentina} is performed to study the reaction $J/\psi\pi^+\pi^-$ in ISR using a data sample of 454 fb$^{-1}$. \\
The $J/\psi\pi^+\pi^-$ mass region below $\sim$4 GeV/c$^2$ is investigated for the first time. 
In that region an excess of events has been observed and the conclusion, after a detailed study of the $\psi(2S)$ lineshape (to estimate the $\psi(2S)$ tail contribution to that region), is that it is not possible to discount the possibilty of a contribution from a $J/\psi\pi^+\pi^-$ continuum cross section in this region. From this study we obtain the cross section value 14.05 $\pm$ 0.26 (stat) pb for radiative return to the $\psi(2S)$ and the measurement of the width $\Gamma(\psi(2S)\rightarrow e^+e^-)=2.31\pm0.05$ (stat) keV. Figure \ref{Fig:vale1}(a) shows the fit to the $J/\psi\pi^+\pi^-$ distribution. A clear signal of the Y(4260) is observed for which the values obtained are $m[Y(4260)]=4244\pm5\pm4$ MeV/c$^2$, $\Gamma[Y(4260)]=114^{+16}_{-15}\pm7$ MeV and $\Gamma_{ee}$ x $\cal{B}$$(J/\psi\pi^+\pi^-)=9.2\pm0.8$ (stat) $\pm$ 0.7 (syst) eV. No evidence for the state at $\sim$ 4 GeV/c$^2$ reported by Belle \cite{Ybelle} was seen. A study of the $\pi^+\pi^-$ system from the Y(4260) decay to $J/\psi\pi^+\pi^-$ is done. The dipion system is in a predominantly S-wave state. The mass distribution exhibits an $f_0(980)$ signal, for  which a simple model indicates a branching ratio with respect to $J/\psi\pi^+\pi^-$ of 0.17 $\pm$ 0.13 (stat). The fit to the dipion invariant mass distribution is shown in Figure \ref{Fig:vale1}(b).

{\raggedright
\begin{footnotesize}

\end{footnotesize}
}

\begin{thebibliography}{99}
\bibitem{belleX3915} S. Uehara {\it et al.} (Belle Collaboration), Phys. Rev. Lett. {\bf 104}, 092001 (2010).
\bibitem{babarX3915B} P. del Amo Sanchez {\it et al.} (\babar Collaboration), Phys. Rev. D. {\bf 82}, 011101(R) (2010).
\bibitem{belleZ} R. Mizuk {\it et al.} (Belle Collaboration), Phys. Rev. D {\bf 78}, 072004 (2008).
\bibitem{antimo} J. P. Lees {\it et al.} (\babar Collaboration), Phys. Rev. D {\bf 85}, 052003 (2012).
\bibitem{Y} B. Aubert {\it et al.} (\babar Collaboration), Phys. Rev. Lett. {\bf 95}, 142001 (2005).
\bibitem{Ybelle} C. Z. Yuan {\it et al.} (Belle Collaboration), Phys. Rev. Lett. {\bf 99}, 182004 (2007).
\bibitem{valentina} \babar Collaboration, \href{http://arxiv.org/abs/1204.2158v1}{{\ttfamily hep-ex/1204.2158v1}}
\end{thebibliography}
\end{document}